\documentclass [12pt,a4paper]{article}
\usepackage{amssymb, theorem}
\newcommand{\<}{\langle}
\renewcommand{\>}{\rangle}

\def\proof{\noindent{\it Proof.} }

\def\bbbr{{\mathbb R}}
\def\bbbc{{\mathbb C}}

\def\ffi{\varphi}

\def\Tr{\mathrm{Tr}\,}

\def\Diag{\mbox{Diag}\,}

\def\ima{\mathrm{Im}}
\def\im{\mathrm{i}}

\def\Det{\mbox{Det}\,}

\def\ccr{{\rm CCR}}
\def\CCR{{\rm CCR}}
\def\fel{\textstyle{\frac{1}{2}}}

\def\iC{\mathcal C}
\def\iH{{\cal H}}
\def\iK{{\cal K}}
\def\iA{{\cal A}}

\def\iF{{\cal F}}

\def\ot{\otimes}
\def\osum{\oplus}
\def\ffi{\varphi}
\newtheorem{thm}{Theorem}[section]
\newtheorem{lemma}{Lemma}[section]
\newtheorem{cor}{Corollary}[section]
\theorembodyfont{\rm}
\newtheorem{example}{Example}[section]
\def\qed{\nobreak\hfill $\square$}
\begin{document}
\ \vskip 1cm 
\centerline{\LARGE {\bf Markov triplets on CCR-algebras}}
\bigskip
\bigskip
\bigskip
\centerline{\large Anna Jen\v cov\'a\footnote{E-mail: jenca@mat.savba.sk. 
Supported by the grants VEGA 2/0032/09 and  APVV- 0071-06.}$^{,4}$,  
D\'enes Petz\footnote{E-mail: petz@math.bme.hu.
Partially supported by the Hungarian Research Grant OTKA  T068258.}$^{,5,6}$
and J\'ozsef Pitrik\footnote{E-mail: pitrik@math.bme.hu. Partially supported 
by the Hungarian Research Grant OTKA TS049835.}$^{,5}$}
\bigskip

\begin{center}
$^{4}$Mathematical Institute of the \\
Slovak Academy of Sciences \\
\v Stef\'anikova 49, 814 73 Bratislava, Slovakia
\end{center}

\begin{center}
$^{5}$Department for Mathematical Analysis, \\
Budapest University of Technology and Economics\\
H-1521 Budapest XI., Hungary
\end{center}

\begin{center}
$^6$ Alfr\'ed R\'enyi Institute of Mathematics, \\
H-1364 Budapest, POB 127, Hungary
\end{center}

\bigskip
\begin{abstract}
The paper contains a detailed computation about the algebra of canonical 
commutation relation, the representation of the Weyl unitaries, the
quasi-free states and their von Neumann entropy. The Markov triplet is
defined by constant entropy increase. The Markov property of a quasi-free
state is described by the representing block matrix. The proof is based
on results on the statistical sufficiency in the non-commutative case.
The relation to classical Gaussian Markov triplets is also described.


\end{abstract}

\medskip
\section*{Introduction}

The notion of quasi-free state was developed in the framework of the 
C*-algebraic approach to the canonical commutation relation (CCR)
\cite{M-V, Holevo, B-R, PD}. The CCR-algebra is generated by the Weyl
unitaries (satisfying a commutation relation, therefore Weyl algebra 
is an alternative terminology). The quasi-free states on CCR can be 
regarded as analogues of Gaussian distributions in classical probability: 
The $n$-point functions can be computed from the 2-point functions and 
in a kind of central limit theorem the limiting state is quasi-free and 
it maximizes the von Neumann entropy when the 2-point function is fixed 
\cite{PDcl}. The quasi-free states are quite tractable, for example 
the von Neumann entropy has an explicit expression \cite{vD-V,B-R}. 

The Markov property was invented by Accardi in the non-commutative
(or quantum probabilistic) setting \cite {Ac0, Ac1, Ac, AcF}. This 
Markov property is based on a completely positive, identity preserving map, 
so-called quasi-conditional expectation and it was formulated in the
tensor product of matrix algebras. A state of a tensor product 
system is Markovian if and only if the von Neumann entropy increase is
constant. This property and a possible definition of the Markov condition
was suggested in \cite{PD94}. A remarkable property of the von Neumann 
entropy is the strong subadditivity \cite{LR, HA, OP, PD08} which plays 
an important role in the investigations of quantum system's correlations. 
The above mentioned constant increase of the von Neumann entropy is
the same as the equality for the strong subadditivity of von Neumann entropy. 

A CCR (or Weyl) algebra is parametrized by a Hilbert space, so we use the
notation $\ccr(\iH)$ when $\iH$ is the Hilbert space. Assume that
$\varphi_{123}$ is a state on the composite system $\ccr(\iH_1) \ot 
\ccr(\iH_2) \ot \ccr(\iH_3)$. Denote by $\varphi_{12}, \varphi_{23}$ the
restrictions to the first two and to the second and third factors, similarly
$\varphi_{2}$ is the restriction to the second factor. The Markov 
property is defined as
$$
S(\varphi_{123})- S(\varphi_{12})=S(\varphi_{23})- S(\varphi_{2}),
$$
where $S$ denotes the von Neumann entropy. When $\varphi_{123}$ is quasi-free,
it is given by a positive operator (corresponding to the 2-point function) 
and the main goal of the present paper is to describe the Markov property 
in terms of this operator. The paper \cite{Pi} studies a similar question 
for the CAR algebra and \cite{APD} is about multivariate Gaussian 
distributions. Although the multivariate Gaussian case (in classical 
probability) is rather different from the present non-commutative setting, 
we use the same block matrix formalism (and the paper \cite{APD} was 
actually a preparation of this problem). The proof of the main result uses
the description of sufficient statistics in the non-commutative case.
A quasi-free state is described by a block matrix and the Markov property 
is formulated by the entries. A Markovian quasi-free state induces 
multivariate Gaussian restrictions, but they are very special in that
framework.

The paper is organized as follows. The preliminary section contains 
some crucial properties of the Weyl unitaries, the Fock space, the CCR 
algebra and quasi-free states. This is written for the sake of completeness,
the results are known but not well-accessible in the literature, cf.
\cite{Exner, B-R, Holevo}. The main
point is the von Neumann entropy formula which is well-known for the CCR 
quasi-free state. In the next section we investigate the quasi-free Markov 
triplets. We obtain a necessary and sufficient condition described in the 
block matrix approach: The block matrix should be block diagonal. There 
are nontrivial Markovian quasi-free states which are not a product in the 
time localization. The existence of such state is interesting, because it 
is in contrast to the CAR case \cite{Pi}. However, the first and the third 
subalgebras are always independent. Finally we prove that commuting field 
operators form a classical Gaussian Markov triplet.

\section{CCR algebras and quasi-free states}
\subsection{Introduction to Weyl unitaries}

In this part the basis of Hermite functions of the Hilbert space $L^2(\bbbr)$
is described in details, the creation, annihilation operators and the Weyl 
unitaries are constructed.

The {\bf Hermite polynomials}
\begin{equation}\label{E:H.1}
H_n(x):=(-1)^n e^{x^2} \frac{d^n}{dx^n}e^{-x^2} \qquad (n=0,1,\dots )
\end{equation}
are orthogonal in the Hilbert space $L^2(\bbbr, e^{-x^2}\,dx)$, they
satisfy the recursion
\begin{equation}
H_{n+1}(x)-2xH_n(x)+2nH_{n-1}(x)=0
\label{E:H.3}
\end{equation}
and the differential equation
\begin{equation}
H_n''(x)-2xH_{n}'(x)+2nH_n(x)=0.         \label{E:H.5}
\end{equation}
The normalized Hermite polynomials
\begin{equation}
\tilde H_n(x)=\frac{1}{\sqrt{2^n n! \sqrt{\pi}}} H_n(x)
\label{E:H.6}
\end{equation}
form an orthonormal basis. From this basis of $L^2(\bbbr, e^{-x^2}\,dx)$,
we can get easily a basis in $L^2(\bbbr)$:
\begin{equation}\label{E:Hfv}
\varphi_n(x):= e^{-x^2/ 2}\tilde H_n(x) \, .
\end{equation}
These are called {\bf Hermite functions}. In terms of the Hermite
functions equation (\ref{E:H.3}) becomes
\begin{equation}
x \varphi_n(x)=\frac{\sqrt{n}\varphi_{n-1}(x)+\sqrt{n+1}\varphi_{n+1}(x)}
{\sqrt{2}} \label{E:H.3B}.
\end{equation}
If the operators $a$ and $a^+$ are defined as
\begin{equation}\label{E:kreal}
a\ffi_n= \sqrt{n} \ffi_{n-1}, \qquad  a^+\ffi_n= \sqrt{n+1} \ffi_{n+1}
\end{equation}
with $a\ffi_0=0$ and the multiplication by the variable $x$ is denoted by $Q$, then
(\ref{E:H.3B}) is
\begin{equation}
Q=\frac{1}{\sqrt{2}}(a+ a^+). \label{E:H.3C}
\end{equation}
From the equation
$$
\frac{\partial}{\partial x} \Big(H_n(x)e^{-x^2/2}\Big)=
H'_n(x)e^{-x^2/2}- x H_n(x)e^{-x^2/2},
$$
one can obtain
\begin{equation}
P\ffi_n :=\frac{1}{\im}
\varphi'_n =\frac{\sqrt{n}\varphi_{n-1}-\sqrt{n+1}\varphi_{n+1}}
{\im\sqrt{2}} \label{E:H.3D},
\end{equation}
that is
\begin{equation}
P=\frac{\im}{\sqrt{2}}(a^+- a). \label{E:H.3E}
\end{equation}
Therefore,
$$
a=\frac{1}{\sqrt{2}}(Q+\im P),\qquad  a^+=\frac{1}{\sqrt{2}}(Q-\im P).
$$

\begin{lemma}
For $z \in \bbbc$ the identity
$$
e(z):=\sum_{n=0}^\infty \frac{z^n}{\sqrt{n!}}\ffi_n(x)=
\pi^{-1/4}\exp \Big( -\frac{z^2+x^2}{2}\Big) \exp (zx \sqrt{2})
$$
holds. Moreover,
$$
e(z)= e^{z a^+}\ffi_0, \qquad \|e(z)\|=e^{|z|^2/2}.
$$
\end{lemma}

\proof 
The identity can be deduced from the generator function
\begin{equation}\label{E:genH}
\sum_{n=0}^\infty \frac{t^n}{n!}H_n(x)=\exp (2xt-t^2)
\end{equation}
of the Hermite polynomials. \qed

The above $e(z)$ is called {\bf exponential vector}.

For $z \in \bbbc$, the operator $\im(za -\bar z a^+)$ is defined
originally on the linear combinations of the basis vectors $\ffi_n$ 
and it is a symmetric operator. It can be proven that its closure 
is self-adjoint, therefore $\exp (za -\bar z a^+)$ becomes a unitary.
\begin{equation}
W(z):= e^{za-\bar z a^+} \label{E:W}
\end{equation} 
is called {\bf Weyl unitary}. Note that
$$
W(z)=\exp \im \sqrt{2}(\alpha P + \beta Q)
$$
if $z=\alpha + \im \beta$. Multiple use of the identity
\begin{equation}
e^{\im (t Q+ u P)}=\exp (\im tu/2) e^{\im t Q}e^{\im u P}=
\exp (-\im tu/2) e^{\im uP}e^{\im tQ}
\end{equation}
gives the following result.

\begin{thm}
$$
W(z)W(z')=W(z+z')\exp ( \im\, \ima (\bar z z'))
$$
for $z,z' \in \bbbc$. 
\end{thm}
With straightforward computation one gets the following.
\begin{lemma}
$$
e^{za-\bar z a^+}\ffi_0=e^{-|z|^2/2} e^{z a^+}\ffi_0=\frac{e(z)}{\|e(z)\|}.
$$
\end{lemma}

The functions
\begin{equation}\label{E:Lag}
L_n^{\alpha}(x)=\sum_{k=0}^n \frac{(-1)^k(n+\alpha)!}{k!(n-k)!(\alpha+k)!}
x^k\qquad(\alpha>-1)
\end{equation}
are called {\bf associated Laguerre polynomials}. We write simply $L_n(x)$ 
for $\alpha=0$. 

\begin{thm}
For $n\ge m$
$$
\< \ffi_m ,W(z)\ffi_n\>=
 e^{-|z|^2/2}\sqrt{\frac{m!}{n!}}z^{n-m} L_m^{n-m}(|z|^2)  
$$
holds.
\end{thm}

\proof
First note that definition (\ref{E:kreal}) implies
\begin{equation}\label{E:A1}
a^k(a^+)^n\ffi_0=
\left\{\begin{array}{ll}
 \frac{n!}{(n-k)!}(a^+)^{n-k}\ffi_0 & \mbox{if $k\le n$,}\\\phantom{MM}\\
 0 & \mbox{if $k>n$.}
 \end{array} \right.
\end{equation}
If $[A,B]$ commutes with $A$ and $B$, then the formula $e^Ae^B=e^{[A,B]/2}
e^{A+B}$ holds. Since $[-\bar{z}a^+,za]=|z|^2I$, we can write
\begin{eqnarray*}
W(z)\ffi_n & = & e^{za-\bar{z}a^+}\ffi_n=e^{-|z|^2/2}
e^{-\bar{z}a^+}e^{za}\ffi_n\\
 & = & \frac{e^{-|z|^2/2}e^{-\bar{z}a^+}}{\sqrt{n!}}
\sum_{k=0}^{\infty}\frac{z^k}{k!}a^k(a^+)^n\ffi_0\\
 & = & \frac{e^{-|z|^2/2}e^{-\bar{z}a^+}}{\sqrt{n!}}
\sum_{k=0}^n\frac{z^kn!}{k!(n-k)!}(a^+)^{n-k}\ffi_0.
\end{eqnarray*}

Now we can compute the matrix elements:
\begin{eqnarray*}
\<\ffi_m, W(z)\ffi_n\> & = & 
\frac{e^{-|z|^2/2}}{\sqrt{m!\,n!}}
\sum_{k=0}^n\sum_{\ell=0}^{\infty}
\frac{(-\bar{z})^\ell z^kn!}{\ell!\,k!\,(n-k)!}
\<(a^+)^m\ffi_0,(a^+)^{n-k+\ell}\ffi_0\>
\\ \phantom{MM} \\
 &=&\frac{e^{-|z|^2/2}}{\sqrt{m!\,n!}}\sum_{k=0}^n\sum_{\ell=0}^{m}
\frac{(-\bar{z})^\ell z^kn!\,m!}{\ell!\,k!\,(n-k)!(m-\ell)!}
\<(a^+)^{m-\ell}\ffi_0,(a^+)^{n-k}\ffi_0\>
\\ \phantom{MM} \\
 & = & e^{-|z|^2/2}\sum_{k=0}^n\sum_{\ell=0}^{m}
\frac{(-\bar{z})^\ell z^k}{\ell!\,k!}
\sqrt{\frac{m!\,n!}{(n-k)!\,(m-\ell)!}}\<\ffi_{m-\ell},\ffi_{n-k}\>
\\ \phantom{MM} \\
 & = & e^{-|z|^2/2}\sum_{k=0}^n\sum_{\ell=0}^{m}
\frac{(-\bar{z})^\ell z^k}{\ell!\,k!}
\sqrt{\frac{m!\,n!}{(n-k)!\,(m-\ell)!}}\delta_{m-l,n-k},
\end{eqnarray*}
where $\delta_{k,\ell}$ denotes the Kronecker symbol. For $n\ge m$, we get 
non-vanishing elements if and only if $k=n-m+\ell$, where $n-m\le k\le n$
and by the formula (\ref{E:Lag}) we obtain
\begin{eqnarray*}
\<\ffi_m,W(z)\ffi_n\> & = & e^{-|z|^2/2}
\sqrt{\frac{m!}{n!}}\sum_{l=0}^m
\frac{(-1)^\ell |z|^{2\ell}z^{n-m}n!}{\ell!(m-l)!(n-m+l)!}\\
& = & e^{-|z|^2/2}\sqrt{\frac{m!}{n!}}z^{n-m} L_m^{n-m}(|z|^2),
\end{eqnarray*}
as we stated. \qed

Note that the case $m\ge n$ can be read out from the Theorem, since
$$
\< \ffi_m ,W(z)\ffi_n\>=\overline{\< \ffi_n ,W(-z)\ffi_m\>}.
$$
The case $m=n$ involves the Laguerre polynomials. The analogue of (\ref{E:genH})
is the formula
\begin{equation}\label{E:genL}
\sum_{n=0}^\infty t^n L_n(x)=\frac{1}{1-t}\exp \Big( -\frac{xt}{1-t}\Big)
\end{equation}
which holds for $|t| <1$ and $x \in \bbbr^+$. This formula is used to obtain
\begin{equation}\label{E:density}
\sum_{n=0}^\infty \mu^n(1-\mu) \< \ffi_n ,W(z)\ffi_n\>= \exp \Big(-\frac{|z|^2}
{2}\frac{1+\mu}{1-\mu}\Big)
\end{equation}
for $0 < \mu <1$. Note that 
\begin{equation}\label{E:D}
D= \sum_{n=0}^\infty \mu^n(1-\mu) |\ffi_n\>\< \ffi_n|
\end{equation}
is a statistical operator (in spectral decomposition). In the corresponding 
state the self-adjoint operator
$$
\frac{za-\bar z a^+}{\im}=(-\im z)a+ \overline{(-\im z)}a^+
$$
has Gaussian distribution.

\subsection{The Fock space}

Let $\iH$ be a Hilbert space. If $\pi$ is a permutation of the numbers
$\{1,2,\dots, n\}$, then on the $n$-fold tensor product $\iH^{\ot n}:=
\iH \ot \iH \ot \dots \ot \iH$ we have a unitary $U_\pi$ such that
$$
U_\pi (f_1 \ot f_2 \ot \dots \ot f_n)=f_{\pi (1)} \ot f_{\pi(2)} \ot \dots 
\ot f_{\pi(n)}.
$$  
The operator
$$
P_n(f_1 \ot f_2 \ot \dots \ot f_n):=\frac{1}{n!}
\sum_{\pi}f_{\pi (1)} \ot f_{\pi(2)} \ot \dots \ot f_{\pi(n)}
$$
is a projection onto the symmetric subspace
$$
\iH^{\vee n}:= \{ g \in \iH^{\ot n}: U_\pi g=g \mbox{\ for\ every\ }\pi\}
$$
which is the linear span of the vectors
$$
|f_1,f_2,\dots, f_n\>\equiv  
f_1 \vee f_2 \vee \ldots \vee f_n: = {1 \over \sqrt{n!}} \sum_{\pi} f_{\pi
(1)} \otimes f_{\pi (2)} \otimes \dots \otimes f_{\pi (n)}\,,
$$
where $f_1 , f_2 , \dots , f_n \in \iH$. Obviously,
$$
f_1 \vee f_2 \vee \ldots \vee f_n= f_{\pi(1)} \vee f_{\pi(2)} \vee 
\ldots \vee f_{\pi(n)}
$$
for any permutation $\pi$.

Assume that  $e_1 , e_2 , \ldots , e_m$ is a basis in $\iH$.
When we consider a vector
$$
e_{i(1)} \vee e_{i(2)} \vee \dots \vee e_{i(n)}
$$
in the symmetric tensor power $\iH^{\vee n}$, we may assume that
$1\le i(1) \le i(2) \le \dots \le i(n) \le m$. A vector $e_t$ may
appear several times, assume that its multiplicity is $r_t$, that is,
$r_t:=\{ \ell: i(\ell)=t\}$. The norm of the vector is $\sqrt{r_1!\,r_2!
\dots r_m!}$ and 
\begin{equation}\label{E:baz}
\Big\{ \frac{1}{\sqrt{r_1!\,r_2!\dots r_m!}}\,
e_{i(1)} \vee e_{i(2)} \vee \dots \vee e_{i(n)} \ : \ 1\le i(1) \le
i(2) \le \dots \le i(n) \le m \Big\}
\end{equation}
is an orthonormal basis in $\iH^{\vee n }$. Another notation is
$$
| e_1^{r_1}, e_2^{r_2}, \dots , e_m^{r_m} \>\equiv 
e_{i(1)} \vee e_{i(2)} \vee \dots \vee e_{i(n)}.
$$

The {\bf symmetric Fock space} is the direct sum
$$
\iF(\iH):=\bbbc \Phi \osum \iH^{\vee 1 }\osum \iH^{\vee 2}\osum \dots
$$
where $\Phi$ is called the vacuum vector and in this spirit the summand 
$\iH^{\vee n}$ is called the $n$-particle subspace. Since $\iH^{\vee 1}$
is identical with $\iH$, the Hilbert space $\iF(\iH)$ is an extension of
$\iH$. The union of the vectors ({\ref{E:baz}) (for every $n$) is a basis 
of the Fock space. 

\begin{lemma} \label{L:1.3}
If $\iH=\iH_1 \osum \iH_2$, then 
$\iF(\iH)=\iF(\iH_1)\ot \iF(\iH_2)$.
\end{lemma}

\proof
It is enough to see that
$$
(\iH_1 \osum \iH_2)^{n \vee}=\iH_1^{\vee n}\osum (\iH_1^{\vee (n-1)} 
\ot \iH_2) \osum \dots (\iH_1 \ot \iH_2^{\vee (n-1)}) \osum \iH_2^{\vee n}.
$$
If $e_1,e_2,\dots, e_m$ is a basis in $\iH_1$ and $f_1,f_2,\dots, f_k$
is a basis in $\iH_2$, then the (non-normalized) basis vector
$$
e_{i(1)} \vee e_{i(2)} \vee \dots \vee e_{i(t)}\vee
f_{j(1)} \vee f_{j(2)} \vee \dots \vee f_{j(n-t)}
$$
can be identified with
$$
e_{i(1)} \vee e_{i(2)} \vee \dots \vee e_{i(t)}\ot
f_{j(1)} \vee f_{j(2)} \vee \dots \vee f_{j(n-t)}
$$
which is a basis vector in $\iH_1^{\vee t}\ot \iH_2^{\vee (n-t)}$.\qed

For $f\in \iH$ the {\bf creation operator} $a^+(f)$ is defined as
\begin{equation}
a^+(f)|f_1 , f_2 , \dots , f_n\>=|f, f_1 , f_2 , \dots , f_n\>.
\end{equation}
$a^+(f)$ is linear in the variable $f$ and it maps the $n$-particle
subspace into the $(n+1)$-particle subspace. Its adjoint is the {\bf 
annihilation operator} which acts as
\begin{equation}
a(f)|f_1 , f_2 , \dots , f_n\>=\sum_{i=1}^n \<f,f_i\>
|f_1 , \dots, f_{i-1}, f_{i+1} , \dots ,f_n\>.
\end{equation}

Given an operator $A\in B(\iH)$ acting on the one-particle space, we
can extend it to the Fock space as follows.
\begin{equation}
\iF(A)|f_1 , f_2 , \dots , f_n\>=\sum_{i=1}^n 
|f_1 , \dots, f_{i-1}, Af_i, f_{i+1} , \dots, f_n\>.
\end{equation}

The next lemma can be shown by simple computation.

\begin{lemma}
For $f,g \in \iH$, we have
$$
\iF(|f\>\<g|)=a^+(f)a(g).
$$
\end{lemma}

Another possibility for extension, or second quantization, of an operator
$U \in B (\iH)$ is given by
\begin{equation}\label{E:Gam}
\Gamma (U) |f_1, f_2, \dots , f_n \> = |U f_1, U f_2, \dots ,U f_n \> .
\end{equation}
It is easy to see that
\begin{lemma}
$$
\Gamma(U_1U_2)=\Gamma (U_1) \Gamma (U_2) \qquad \mbox{and} \qquad
\Gamma(U^*)=\Gamma(U)^*.
$$
\end{lemma}

If $U$ is a unitary, then $\Gamma (U)$ is unitary as well. Moreover,
if $U(t)$ is a continuous one-parameter group with generator $A$, then 
the generator of the continuous one-parameter group $\Gamma (U (t))$
on $\iF(\iH)$ is the closure of $\iF(A)$. To show an example, we note that 
the statistical operator (\ref{E:D}) is $(1-\mu)\Gamma (\mu)$ in
the case of a one-dimensional $\iH$. 

\begin{lemma}
Let  $\iH=\iH_1 \osum \iH_2$ and $U=U_1 \osum U_2$. Then
$$
\Gamma(U_1\osum U_2)=\Gamma (U_1) \ot \Gamma (U_2).
$$
\end{lemma}


\subsection{The algebra of the canonical commutation relation}

Let $\iH$ be a finite-dimensional Hilbert space. 
Assume that for every $f \in \iH$ a unitary
operator $W(f)$ is given such that the relations
\begin{equation}\label{E:W2}
W(f_1)W(f_2)=W(f_1+f_2)\exp ( \im\, \sigma ( f_1, f_2))
\end{equation}
\begin{equation}\label{E:W2b}
W(-f)=W(f)^*
\end{equation}
hold for $f_1,f_2,f \in \iH$ with $\sigma ( f_1, f_2):=\ima \<f_1, f_2\>$.
The abstract C*-algebra generated by these unitaries is unique and denoted
by $\ccr(\iH)$. The relation (\ref{E:W2}) shows that $W(f_1)$ and $W(f_2)$
commute if $f_1$ and $f_2$ are orthogonal. Therefore for an $n$-dimensional
$\iH$, the algebra $\ccr(\iH)$ is an $n$-fold tensor product 
$$
\ccr(\bbbc) \ot \dots \ot \ccr(\bbbc).
$$

Since $W(tf)W(sf)=W((t+s)f)$ for $t,s \in \bbbr$, the mapping $t 
\mapsto W(tf)$ is
a one-parameter unitary group which is not norm continuous since 
$\|W(f_1)-W(f_2)\| \ge \sqrt{2}$ when $f_1 \ne f_2$ \cite{PD}.

The C*-algebra $\ccr(\iH)$ has a very natural state 
\begin{equation}\label{E:Fstate}
\omega (W(f)):= \exp \big(- \|f\|^2/2\big)
\end{equation}
which is called {\bf Fock state}. The GNS-representation of $\ccr(\iH)$
is called {\bf Fock representation} and it leads to the the Fock space
$\iF(\iH)$ with cyclic vector $\Phi$. If  $f_1$ and $f_2$ are orthogonal
vectors, then
\begin{eqnarray*}
\omega(W(f_1)W(f_2))&=&\omega(W(f_1+f_2))= \exp \big(- \|f_1+f_2\|^2/2\big)
\cr &=&\exp \big(- \|f_1|^2/2\big)\exp \big(- \|f_2\|^2/2\big)=\omega(W(f_1))
\omega(W(f_2)).
\end{eqnarray*}
Therefore $\omega$ is a product state and it follows that the GNS Hilbert 
space is a tensor product. (This is another argument to
justify Lemma \ref{L:1.3}.) We shall identify the abstract
unitary $W(f)$ with the representing unitary acting on the tensor
product GNS-space $\iF(\iH)$. The map 
$$t\mapsto \pi_\Phi(W(tf))$$
is an so-continuous 1-parameter group of unitaries, and according to 
the Stone theorem
$$
\pi_\Phi(W(tf))=\exp(\im tB(f))
$$
for a self-adjoint operator $B(f)$, called {\bf field operator}.
Let
$$
B^{\pm}(f)=\frac{1}{2}(B(f)\mp\im B(\im f)).
$$
Then 
$$
[B^-(f),B^+(g)]=\<g,f\>
$$
is the canonical commutation relation for the {\bf creation operator} $B^+(g)$ 
and the {\bf annihilation operator} $B^-(f)$.
When $\iH=\bbbc$, then
$$
W(z)= \exp \im(a(z)+a^+(z)),
$$
where $ a^+(z)=\im\bar{z}a^+ $.

\subsection{Quasi-free states}

The Fock state (\ref{E:Fstate}) can be generalized by choosing a positive
operator $A \in B(\iH)$:
\begin{equation}\label{E:Fstate2}
\omega_A (W(f)):= \exp \big(- \|f\|^2/2 -\< f, Af\> \big).
\end{equation}
This is called {\bf quasi-free state} \cite{M-V}.

\begin{example}
Assume that $\iH=\iH_1 \osum \iH_2$ and write the positive mapping
$A \in B(\iH)$ in the form of block matrix:
$$
A=\left[\matrix{ A_{11} & A_{12} \cr A_{21} & A_{22} }\right].
$$
If $f \in \iH_1$, then
$$
\omega_A (W(f \osum 0))= \exp \big(- \|f\|^2/2 -\< f, A_{11}f\> \big).
$$
Therefore the restriction of the quasi-free state $\omega_A$ to
$\ccr(\iH_1)$ is the quasi-free state $\omega_{A_{11}}$. 

If
$$
A=\left[\matrix{ A_{11} & 0 \cr 0 & A_{22} }\right],
$$
then $\omega_A=\omega_{A_{11}} \ot \omega_{A_{22}}$.  \qed 
\end{example}

\begin{example}
Assume that $\iH$ is one-dimensional and let $A=\lambda>0$. We can read out
from formulas (\ref{E:density}) and (\ref{E:D}) that the statistical
operator of $\omega_\lambda$ in the Fock representation is
\begin{equation}\label{E:lD}
D_\lambda= \sum_{n=0}^\infty \frac{1}
{1+\lambda}\Big(\frac{\lambda}{1+\lambda}\Big)^n |\ffi_n\>\< \ffi_n|.
\end{equation}
(Note the $\mu=\lambda/(1+\lambda)$ in (\ref{E:D}).) Moreover,
\begin{equation}\label{E:2point}
\omega_\lambda(a^+a)=\lambda.
\end{equation}
One can easily compute the von Neumann entropy of the state $\omega_\lambda$
from the eigenvalues of the statistical operator $D_\lambda$:
\begin{equation}\label{E:ent}
S(\omega_\lambda)=\eta(\lambda)- \eta(\lambda+1),
\end{equation}
where $\eta(\lambda)=- \lambda \log \lambda$.

The case of finite-dimensional $\iH$ can be reduced to the one-dimensional by
the spectral decomposition of the operator $A$. \qed
\end{example}

Assume that $\omega$ is a state of $\ccr(\iH)$. If
$$
C_\omega (f,g):= \omega(a^+(f)a(g))
$$
can be defined, then it will be called {\bf 2-point function} of $\omega$.

\begin{thm}
Assume that the spectral decomposition of $0 \le A \in B(\iH)$ is
\begin{equation}\label{E:specA}
A=\sum_{i=1}^m \lambda_i |e_i\>\< e_i|.
\end{equation}
Then the statistical operator
of the quasi-free state $\omega_A$ in the Fock representation is
\begin{equation}\label{E:denA}
D_A= \left(\prod_{i=1}^m  \frac{1}{1+\lambda_i}\right)\sum_{r_j}
\left(\prod_{i=1}^m
\Big(\frac{\lambda_i}{1+\lambda_i}\Big)^{r_i}\frac{1}{r_i!}\right)
| e_1^{r_1}, e_2^{r_2}, \dots , e_m^{r_m} \>
\<e_1^{r_1}, e_2^{r_2}, \dots , e_m^{r_m} |,
\end{equation}
where summation is over $n=0,1,2 \dots$ and the decompositions
$n=r_1+r_2+\dots +r_m$. Moreover,
$$
\omega_A (a^+(f)a(g))=\<g, Af\> \qquad (f,g \in \iH)
$$
and 
$$
S(\omega_A)=\Tr \eta(A)- \Tr \eta(A+I).
$$
\end{thm}

\proof
The basic idea is the decomposition
\begin{equation}\label{E:Aprod}
\omega_A = \omega_{\lambda_1}\ot \omega_{\lambda_2}\ot 
\dots \ot \omega_{\lambda_m}
\end{equation}
when the space $\iH$ is decomposed into the direct sum of the one-dimensional
subspaces $\bbbc |e_i\>$ and $\iF(\iH)$ and $\ccr(\iH)$ become tensor product.
The statistical operator of $\omega_{\lambda_i}$ is
$$
D_{\lambda_i}= \sum_{r_i=0}^\infty \frac{1}{1+\lambda_i}
\Big(\frac{\lambda_i}{1+\lambda_i}\Big)^{r_i}\frac{1}{r_i!} 
|e_i^{r_i}\>\< e_i^{r_i}|
$$
the tensor product is exactly the stated matrix.

When we want to check the 2-point function, it is enough to consider
the case $f=g=e_i$. This is OK due to (\ref{E:2point}). 

The von Neumann entropy is deduced from (\ref{E:ent}) and (\ref{E:Aprod}).
\qed

If (\ref{E:specA}) holds, then 
$$
\Gamma( A(I+A)^{-1})| e_1^{r_1}, e_2^{r_2}, \dots , e_m^{r_m} \>=
\prod_{i=1}^m \left(\frac{\lambda_i}{1+\lambda_i}\right)^{r_i}
| e_1^{r_1}, e_2^{r_2}, \dots , e_m^{r_m} \>
$$
and we have
\begin{equation}\label{E:DAg}
D_A= \frac{1}{c_A}\Gamma(A(I+A)^{-1})\,,\quad \mbox{where}\quad 
c_A=\Tr\Gamma(A(I+A)^{-1}).
\end{equation}
This leads to the following result.

\begin{thm}\label{T:Con}
Let $\omega_A$ and $\omega_B$ be quasi-free state of $\ccr(\iH)$ 
which correspond to the operators $0 \le A,B \in B(\iH)$. Their
Connes cocycle is
\begin{equation}\label{E:DAg2}
[D \omega_A,D \omega_B]_t=u_t \Gamma\Big((A(I+A)^{-1})^{\im t}
(B(I+B)^{-1})^{-\im t}\Big)
\end{equation}
where 
$$
u_t=\Big( \Tr\Gamma(A(I+A)^{-1})\Big)^{- \im t}
\Big(\Tr\Gamma(B(I+B)^{-1})\Big)^{\im t}.
$$
\end{thm}

\begin{thm}
Let $\omega$ be a state of $\ccr(\iH)$ such that its 2-point function
is $\omega(a^+(f)a(g))=\<g, Af\>$  ($f,g \in \iH$) for a positive
operator $A \in B(\iH)$. Then $S(\omega)\le S(\omega_A)$ and equality
implies $\omega=\omega_A$. 
\end{thm}

\proof
Consider the one-dimensional case when $A=\lambda$. We compute
the relative entropy $S(\omega || \omega_\lambda)$:
\begin{eqnarray*}
S(\omega || \omega_\lambda)
&=&
-S(\omega)-\omega(\log D_\lambda)
\cr &=&
-S(\omega)-\log (1+\lambda)\,\omega\Big(\sum_{n=0}^\infty 
|\ffi_n\>\<\ffi_n|\Big)
-\log \frac{\lambda}{1+\lambda}\,\omega\Big(\sum_{n=0}^\infty 
n |\ffi_n\>\<\ffi_n|\Big)\cr &=&
-S(\omega)- \log (1+\lambda)-\lambda \log \frac{\lambda}{1+\lambda}
=-S(\omega)+S(\omega_\lambda)
\end{eqnarray*}
Since the relative entropy $S(\omega || \omega_\lambda)> 0$ if
$\omega$ and $\omega_\lambda$ are different, the statement is obtained.

The general case can be proved by similar computation. The result was also
obtained in connection with the central limit theorem \cite{PDcl}. \qed

\section{Markov triplets}

Let $\iH=\iH_1 \oplus \iH_2 \oplus \iH_3$ be a finite-dimensional 
Hilbert space and consider the Fock representation of $\ccr(\iH)
\equiv \ccr(\iH_1 \oplus \iH_2 \oplus \iH_3)$ on $\iF(\iH_1 \oplus 
\iH_2 \oplus \iH_3)$. Instead of the C*-algebra, we work with the 
weak closure in the Fock representation: $\iA_{123}:=B(\iF(\iH_1
\oplus \iH_2 \oplus \iH_3)) \equiv B(\iF(\iH_1))\ot B(\iF(\iH_2)) \ot 
B(\iF(\iH_3)))$. This algebra has subalgebras 
\begin{eqnarray*}
\iA_{12}:&=&B(\iF(\iH_1 \oplus \iH_2 \oplus 0)) \equiv
B(\iF(\iH_1))\ot B(\iF(\iH_2)) \ot \bbbc I, \cr
\iA_{23}:&=&B(\iF(0 \oplus \iH_2 \oplus \iH_3)) \equiv
\bbbc I \ot B(\iF(\iH_2))\ot B(\iF(\iH_3)), \cr
\iA_{2}:&=&B(\iF(0 \oplus \iH_2 \oplus 0)) \equiv
\bbbc I \ot B(\iF(\iH_2))\ot \bbbc I.
\end{eqnarray*}

Assume that $D_{123}$ is a statistical operator in $\iA_{123}$ and
we denote by $D_{12}, D_2, D_{23}$ its reductions in the subalgebras
$\iA_{12}, \iA_2, \iA_{23}$, respectively. These subalgebras form a 
{\bf Markov triplet} with respect to the state $D_{123}$ if
\begin{equation}\label{E:tripl}
S(D_{123})-S(D_{23}) =S(D_{12})-S(D_{2}),
\end{equation}
where $S$ denotes the von Neumann entropy and we assume that both sides
are finite in the equation. The state $\omega$ corresponding to the
statistical operator  $D_{123}$ is called {\bf Markov state}.

Condition (\ref{E:tripl}) is equivalent to several other conditions, see, for 
example, Chapter 9 of \cite{PD08} about the details and proofs. In most studies 
about the strong subadditivity of the von Neumann entropy and the equality case
(\ref{E:tripl}), the Hilbert space is assumed to be finite-dimensional. In our setting
the Fock space is always infinite-dimensional, so \cite{JPD} might be the optimal 
reference. Here we prefer equivalent formulation in terms of Connes cocyles:
\begin{equation}\label{E:tripl2}
[D \omega_{123}, D(\ffi_1 \ot \omega_{12})]_t= 
[D(\omega_{12}\ot \ffi_3), D(\ffi_1 \ot \omega_2 \ot \ffi_3)]_t
\end{equation}
for every real $t$, where $\ffi_1$ and $\ffi_3$ are arbitrary states.

Let 
\begin{equation}\label{E:Aop}
A=\left[\matrix{ A_{11} & A_{12} & A_{13}\cr A_{21} & A_{22} & A_{23}
\cr A_{31} & A_{32} & A_{33} }\right].
\end{equation}
be a positive operator on $\iH_1 \oplus \iH_2 \oplus \iH_3$ and assume that
$D_{123}$ is the statistical operator of the quasi-free state $\omega_A$.
$\omega_A$ restricted to $\iA_{23}$ is a quasi-free state 
induced by the operator
$$
\left[\matrix{ A_{22} & A_{23} \cr A_{32} & A_{33}}\right].
$$
Set
$$
D=\left[\matrix{ I & 0 & 0 \cr 0 & A_{22} & A_{23}
\cr 0 & A_{32} & A_{33} }\right], \qquad
B=\left[\matrix{ A_{11} & A_{12} & 0 \cr A_{21} & A_{22} & 0
\cr 0 & 0 & I }\right], \qquad
C=\left[\matrix{ I & 0 & 0 \cr 0 & A_{22} & 0
\cr 0 & 0 & I }\right].
$$
Then equality (\ref{E:tripl2}) may become
\begin{equation}\label{E:tripl12}
[D \omega_{A}, D \omega_{D}]_t= 
[D \omega_{B}, D\omega_C]_t.
\end{equation}
According to Theorem \ref{T:Con} this is the condition
$$
\Gamma\Big((A(I+A)^{-1})^{\im t}(D(I+D)^{-1})^{-\im t}\Big)=\lambda_t
\Gamma\Big((B(I+B)^{-1})^{\im t}(C(I+c)^{-1})^{-\im t}\Big)
$$ 
with a set of numbers $\lambda_t$.

One can see from formula (\ref{E:Gam}) that $\Gamma(U)=\lambda\Gamma(V)$
for a $\lambda \in \bbbc$ implies $\lambda=1$ and $U=V$. Therefore
condition (\ref{E:tripl2}) becomes the following.

\begin{thm}
For a quasi-free state $\omega_A$ the Markov property (\ref{E:tripl}) is 
equivalent to the condition
\begin{equation}\label{E:bmc}
A^{\im t}(I+A)^{-\im t}D^{-\im t}(I+D)^{\im t}=
B^{\im t}(I+B)^{-\im t}C^{-\im t}(I+C)^{\im t}
\end{equation}
for every real $t$.
\end{thm}

The problem is the solution of this equation. 
Note that if condition (\ref{E:bmc}) holds for every real $t$, 
then analytic continuation gives all complex $t$.

\begin{cor}
If $A$ gives a Markov triplet, then $U^*AU$ gives a Markov triplet
as well when $U=\Diag(U_1,U_2,U_3)$ with unitaries $U_i \in B(\iH_i)$,
$1 \le i \le 3$.
\end{cor}

\begin{example}\label{Ex:igaz}
The following matrix satisfies condition (\ref{E:bmc}).
\begin{equation}\label{E:trivi}
A=\left[\matrix{ A_{11} &
\left[\matrix{a & 0}\right]&
0
\cr 
&&\cr
\left[\matrix{a^* \cr 0}\right]&
\left[\matrix{c & 0 \cr 0 & d}\right]&
\left[\matrix{0 \cr b }\right] \cr &&\cr
 0 &
\left[\matrix{0 & b^*}\right]&
A_{33}
 }\right]=\left[\matrix{
\left[\matrix{ A_{11} & a \cr a^* & c}\right] &
0 \cr & \cr & \cr
0 &
\left[\matrix{d & b \cr b^* & A_{33}}\right]
 }\right], 
\end{equation}
where the parameters $a, b, c, d$ (and $0$) are matrices. This 
is a block diagonal matrix, $A=\Diag ( A_1 , A_2 )$, so
we have $f(A)=\Diag ( f(A_1 ), f(A_2 ) )$ for a function $f$. The
matrices $B$, $C$ and $D$ are block diagonal as well:
\begin{eqnarray*}
B&=&\Diag\big( A_1 , \Diag(d, I)\big), \cr
C&=&\Diag\big( \Diag (I, c), \Diag(d, I)\big),\cr 
D&=&\Diag\big( \Diag (I, c), A_2  \big).
\end{eqnarray*}
Therefore,
$$ 
f(A)g(D)=\Diag\big( f(A_1 ) \Diag (g(I), g(c)), f(A_2 ) g(A_2 )\big)
$$
and
$$
f(B)g(C)=\Diag\big( f(A_1 )\Diag (g(I), g(c)), \Diag(f(d)g(d), 
f(I)g(I))\big).
$$
If $fg=1$, then $f(A)g(D)=f(B)g(C)$. 

Note that 
$$
A=\left[\matrix{ A_{11} & 0 &  0\cr 0 & A_{22} & A_{23}
\cr 0 & A_{32} & A_{33} }\right] \mbox{\ and\ }
A=\left[\matrix{ A_{11} & A_{12} & 0 \cr A_{21} & A_{22} & 0
\cr 0 & 0 & A_{33} }\right]
$$
are particular cases. 

On the basis of the previous Corollary we can use  block diagonal unitaries
to have further examples. \qed
\end{example}

\begin{thm}\label{T:2.2}
The condition 
\begin{equation}\label{E:bmc2}
A^{-1}(I+A)D(I+D)^{-1}=
B^{-1}(I+B)C(I+C)^{-1}
\end{equation}
implies that
\begin{equation}\label{E:felk}
A_{13}=A_{12} A_{22}^{-1} A_{23}\quad \mbox{and}
\quad A_{13}=A_{12} (A_{22}+I)^{-1} A_{23}.
\end{equation}
\end{thm}

\proof
In the computation of the inverse of a block matrix, 
the following formula is very useful. 
If $P$ and $S$ are square matrices and $S$ is invertible, then
\begin{equation}\label{E:Schur}
M^{-1}=\left[
\begin{array}{cc}
P & Q\\
R & S
\end{array}
\right]^{-1}=\\
\left[
\begin{array}{cc}
 (M/S)^{-1} & -(M/S)^{-1}QS^{-1}\\
-S^{-1}R(M/S)^{-1} & S^{-1}+S^{-1}R(M/S)^{-1}QS^{-1}
\end{array}
\right],
\end{equation}
where $(M/S)\equiv P-QS^{-1}R$ is the Schur complement of $S$ in $M$ 
(see \cite[Section 7.7]{HJ}, actually, the checking is a simple 
multiplication). If $P$ is also invertible, the equation
\begin{equation}
S^{-1}+S^{-1}R(M/S)^{-1}QS^{-1}=(M/P)^{-1}
\end{equation}
also holds, where $(M/P)\equiv S-RP^{-1}Q$ is the Schur complement of 
$P$ in $M$.
For solving (\ref{E:bmc2}), we partition
the block matrix $A$ in the following way
$$
A=\left[\matrix{ A_{11} & A_{12} & A_{13}\cr A_{21} & A_{22} & A_{23}
\cr A_{31} & A_{32} & A_{33} }\right]=
\left[\matrix{ P & Q \cr Q^* & S}\right],
$$
where we used the fact that $A$ is positive self-adjoint and used the notations
$$
S=\left[\matrix{ A_{22} & A_{23} \cr A_{23}^* & A_{33} }\right],
$$
$P=A_{11}$ and $Q=\left[A_{12} \quad A_{13}\right]$.
With the help of (\ref{E:Schur}) we get
\begin{equation}
A^{-1}(I+A)=
\left[
\begin{array}{cc}
I+(A/S)^{-1} & -(A/S)^{-1}QS^{-1}\\
-S^{-1}Q^*(A/S)^{-1} & I+(A/P)^{-1}
\end{array}
\right].
\end{equation}
Similarly we write $D$ in $2 \times 2$ matrix form:
$$
D=\left[\matrix{ I & 0 & 0 \cr 0 & A_{22} & A_{23}
\cr 0 & A_{32} & A_{33} }\right]=
\left[\matrix{ I & 0 \cr 0 & S}\right],
$$
where $0$ denotes the block matrix $[0\quad 0]$, or its transpose.
Now we can compute the left-hand side of (\ref{E:bmc2}):
\begin{equation}\label{E:LHS}
A^{-1}(I+A)D(I+D)^{-1}=
\left[
\begin{array}{cc}
\frac{1}{2}\left[I+(A/S)^{-1}\right] & -(A/S)^{-1}Q(I+S)^{-1}\\
-\frac{1}{2}S^{-1}Q^*(A/S)^{-1} & \left[I+(A/P)^{-1}\right]S(I+S)^{-1}
\end{array}
\right].
\end{equation}
With a similar procedure we have
$$
B=\left[\matrix{ A_{11} & A_{12} & 0 \cr A_{21} & A_{22} & 0
\cr 0 & 0 & I }\right]=
\left[\matrix{ P & \tilde{Q} \cr \tilde{Q}^* & \tilde{S}}\right],
$$
and
$$
C=\left[\matrix{ I & 0 & 0 \cr 0 & A_{22} & 0
\cr 0 & 0 & I }\right]=
\left[\matrix{ I & 0 \cr 0 & \tilde{S}}\right],
$$
where 
$$
\tilde{S}=\left[\matrix{ A_{22} & 0 \cr 0 & I }\right],
\qquad \tilde{Q}=\left[A_{12} \quad 0\right],
$$
and the same remark concern for the $0$ block matrices as above.
We get for the right-hand side of (\ref{E:bmc2}):
\begin{equation}\label{E:RHS}
B^{-1}(I+B)C(I+C)^{-1}=
\left[
\begin{array}{cc}
\frac{1}{2}\left[I+(B/\tilde{S})^{-1}\right] & 
-(B/\tilde{S})^{-1}\tilde{Q}(I+\tilde{S})^{-1}\\
-\frac{1}{2}\tilde{S}^{-1}\tilde{Q}^*(B/\tilde{S})^{-1} & 
\left[I+(B/P)^{-1}\right]\tilde{S}(I+\tilde{S})^{-1}
\end{array}\right].
\end{equation}
From the equality between (\ref{E:LHS}) and (\ref{E:RHS}) we have equations 
for the block matrices. The equality of $(1,1)$ elements implies $(A/S)=
(B/\tilde{S})$. This and the equality of $(1,2)$ elements gives the 
equation $Q(I+S)^{-1}=\tilde{Q}(I+\tilde{S})^{-1}$, which lead us to 
$A_{13}=A_{12} (A_{22}+I)^{-1} A_{23}$. From the $(2,1)$ elements we have
$S^{-1}Q^*=\tilde{S}^{-1}\tilde{Q}^*$, this implies the other 
necessary condition. The $(2,2)$ elements will be equal automatically 
when these conditions hold.\qed

According to \cite{APD}, (\ref{E:felk}) means that the $(1,3)$ element of $A^{-1}$
and $(A+I)^{-1}$ are 0. It is interesting that if we take the determinant
of equation (\ref{E:bmc}), then we have
\begin{eqnarray*}
&&(\Det A)(\Det C) (\Det D)^{-1} (\Det B)^{-1}\cr && \qquad
=(\Det (I+A))(\Det (I+C))(\Det (I+D))^{-1}  (\Det (I+B))^{-1}.
\end{eqnarray*}
According to Theorem 5 in \cite{APD}, both sides are smaller or equal 
than 1 and (\ref{E:felk}) is equivalent to the condition that both sides 
are exactly 1. 

Let $X$ be the inverse of the block matrix (\ref{E:Aop}) and suppose 
that (\ref{E:bmc2}) holds. Tedious computation yields that
\begin{eqnarray*}
X_{11}&=&\left(A_{11}-A_{12}A_{22}^{-1}A_{21}\right)^{-1},
\cr
X_{12}&=& -\left(A_{11}-A_{12}A_{22}^{-1}A_{21}\right)^{-1}A_{12}A_{22}^{-1},
\cr
X_{13}&=&0,
\cr
X_{22}&=&\left(A_{22}-A_{21}A_{11}^{-1}A_{12}\right)^{-1}+A_{22}^{-1}A_{23}
\left(A_{33}-A_{32} A_{22}^{-1}A_{23}\right)^{-1}A_{32}A_{22}^{-1},
\cr
X_{23}&=&-A_{22}^{-1}A_{23}\left(A_{33}-A_{32} A_{22}^{-1}A_{23}\right)^{-1},
\cr
X_{33}&=&\left(A_{33}-A_{32}A_{22}^{-1}A_{23}\right)^{-1}\,.
\end{eqnarray*}

The next example shows that conditions (\ref{E:felk}) are not sufficient,
in contrast to the classical Gaussian Markov triplets \cite{APD}.

\begin{example}
The matrix
$$
A=\left[\matrix{ \left[\matrix{4 & 0 \cr 0 & 5}\right]&
\left[\matrix{1 & 1 \cr -\frac{2}{7} & -\frac{2}{7}}\right]&
\left[\matrix{\frac{1}{14} & \frac{1}{14} \cr -\frac{1}{49} & -\frac{1}{49}}\right]
\cr 
&&\cr
\left[\matrix{1 & -\frac{2}{7} \cr 1 & -\frac{2}{7}}\right]&
\left[\matrix{6 & 0 \cr 0 & 3}\right]&
\left[\matrix{1 & 1 \cr -\frac{2}{7} & -\frac{2}{7}}\right] \cr &&\cr
\left[\matrix{\frac{1}{14} & -\frac{1}{49} \cr \frac{1}{14} & -\frac{1}{49}}\right]&
\left[\matrix{1 & -\frac{2}{7} \cr 1 & -\frac{2}{7}}\right]&
\left[\matrix{3 & 0 \cr 0 & 1}\right]
 }\right] 
$$
is positive and fulfills conditions (\ref{E:felk}), but (\ref{E:bmc}) does not hold.
Indeed, numerical computation shows that 
$$
\log A(I+A)^{-1}+\log C(I+C)^{-1} \ne \log B(I+B)^{-1}+\log D(I+D)^{-1},
$$
or an alternative argument is that the matrix is different from (\ref{E:trivi}),
cf. Theorem \ref{thm:markov}.

This example shows that condition
\begin{equation}\label{E:tripl33}
D_{123} D_{23}^{-1}= D_{12} D_2^{-1}
\end{equation}
is weaker than 
\begin{equation}\label{E:tripl3}
D_{123}^{\im t}D_{23}^{-\im t}= D_{12}^{\im t} D_2^{-\im t} 
\qquad (t \in \bbbr).
\end{equation}
Note that in the finite-dimensional case (\ref{E:tripl3}) is equivalent
to
$$
D_{123}^{1/2}D_{23}^{-1/2}= D_{12}^{1/2} D_2^{-1/2}, 
$$
see Chapter 9 of \cite{PD08}. \qed
\end{example}

In the notation
$$
K:=A^{-1}(I+A),\ L:=D^{-1}(I+D),\ M:=B^{-1}(I+B),\ N:=C^{-1}(I+C)
$$
condition (\ref{E:bmc}) becomes
\begin{equation}\label{eq:markov}
K^{-\im t}L^{\im t}=M^{-\im t}N^{\im t}.
\end{equation}

\begin{thm}\label{thm:markov} 
The Markov property (\ref{E:tripl}) is satisfied if and only if 
there is a projection $P \in B(\iH)$ such that $P|\iH_1\equiv I$,
$P|\iH_3\equiv 0$ and $PA=AP$. In other words, $A$ is block diagonal
in the form (\ref{E:trivi}).
\end{thm}

\proof
We write the matrices of the relation (\ref{eq:markov}) in block form: 
\begin{eqnarray*}
K&=& \left[\begin{array}{ccc} K_{11} & K_{12} &K_{13} \\
                          K_{21} & K_{22} & K_{23}\\
			  K_{31} & K_{32} & K_{33}
			  \end{array} \right],\quad
L=\left[\begin{array}{ccc} 2 & 0 &0 \\
                          0 & L_{22} & L_{23}\\
			  0 & L_{32} & L_{33}
			  \end{array} \right],\\
M&=&\left[\begin{array}{ccc} M_{11} & M_{12} &0 \\
                          M_{21} & M_{22} & 0\\
			  0 & 0 & 2
			  \end{array} \right],\qquad
N=\left[\begin{array}{ccc} 2 & 0 &0 \\
                          0 & N_{22} & 0\\
			  0 & 0 & 2
			  \end{array} \right].
\end{eqnarray*}

Suppose that the Markov property is satisfied, and we use it in the form 
(\ref{eq:markov}). Since $K=I+A^{-1}$, the block diagonal structure
(\ref{E:trivi}) of $A$ is equivalent to that property of $K$. We shall
work on $K$.

Let $\iC$ be the subalgebra generated by the set 
$\{K^{\im t}L^{-\im t}: t\in \bbbr\}$. By the factorization result 
in \cite{JPD}, there are positive matrices $\tilde X,\tilde Y\in
\iC$ and $0\le \tilde Z\in B(\iH_1\oplus\iH_2\oplus \iH_3)$, such that 
\begin{equation} \label{E:has1}
K=\tilde X\tilde Z, \qquad L=\tilde Y\tilde Z, \qquad \tilde Z\tilde X=\tilde X 
\tilde Z, 
\qquad \tilde Z \tilde Y=\tilde Y \tilde Z.
\end{equation} 
Since (\ref{eq:markov}) implies that $\iC\subseteq B(\iH_1\oplus \iH_2)\oplus 
\bbbc I$, we have  $\tilde X$ and $\tilde Y$ in the above form
$$
\tilde X:=\left[\begin{array}{cc} X & 0\\
                        0 & I
	\end{array}\right] 		
\quad \mbox{and} \quad
\tilde Y:=\left[\begin{array}{cc} Y & 0\\
                        0 & I
	\end{array}\right].
$$ 
We write $\tilde Z$ in a similar block form
\[
\tilde Z=\left[\begin{array}{cc} Z & z\\
                           z^* & \tilde Z_{33}
	\end{array}\right],		   
\]
where $Z\in B(\iH_1\oplus \iH_2)$ and $z^*=[\tilde Z_{31}, \tilde Z_{32}]$. Then
\begin{equation}\label{eq:K1}
K=\tilde X\tilde Z=\left[\begin{array}{cc} XZ & Xz\\
                             z^* & \tilde Z_{33}
	\end{array}\right],
\quad \mbox{and} \quad
L=\tilde Y\tilde Z=\left[\begin{array}{cc} YZ & Yz\\
                             z^* & \tilde Z_{33}
	\end{array}\right].		     
\end{equation}
This implies that 
\begin{equation}\label{eq:XYz}
Xz=Yz=z=\left[\begin{array}{c} 0 \\ L_{23}\end{array}\right],
\end{equation}
$Z$ commutes with $X$ and $Y$ and 
\[
[K_{31},K_{32},K_{33}]= [z^*,\tilde Z_{33}]=
[0,L_{32},L_{33}].
\]
In particular, $K_{31}=K_{13}=0$ and $K_{23}=L_{23}$.

By (\ref{eq:K1}) and (\ref{eq:XYz}), we have
\[
\left[\begin{array}{cc} K_{11} & K_{12} \\
                            K_{21} & K_{22}
		\end{array}\right]\left[\begin{array}{c} 0 \\ L_{23}
		\end{array}\right]=XZz=ZYz=
		\left[\begin{array}{cc} 2 & 0 \\ 0 & L_{22}
		\end{array}\right]z=\left[\begin{array}{c} 0 \\ L_{22}L_{23}
		\end{array}\right]
\]
and we get $K_{12}K_{23}=K_{12}L_{23}=0$.

If the range of $L_{23}$ is $\iH_2$, then $K_{12}=0$, and if $L_{23}=0$, 
then $K_{23}=0$, so in both cases $K$ is block diagonal.

Suppose now that the range of $L_{23}$ is not $\iH_2$ and $L_{23}\ne 0$.
Then there is a decomposition $\iH_2=\iK_a\oplus \iK_b$, where 
$\iK_b$ is the range of $L_{23}$. Next we work in the frame of the
decomposition $(\iH_1 \oplus \iK_a)\oplus \iK_b $.

For each vector $\xi\in \iK_b$, we have $X\xi=Y\xi=\xi$. It follows 
that there are matrices $X_1,Y_1\in B(\iH_1\oplus \iK_a)$, such that
\[
X=\left[\begin{array}{cc} X_1 & 0\\
                          0 & I
	\end{array}\right],\quad  Y=\left[\begin{array}{cc} Y_1 & 0\\
                          0 & I
	\end{array}\right]		  
\]
If we write 
$$
Z=\left[\begin{array}{cc} Z_1 & z_1\\
                          z_1^* & Z_{33}
	\end{array}\right], \qquad Z_1\in B(\iH_1\oplus \iK_a)
\mbox{\ and\ } Z_{33} \in B(\iK_b),
$$
then
\[
XZ=\left[\begin{array}{cc} X_1Z_1 & X_1z_1\\
                          z_1^* & Z_{33}
	\end{array}\right]=\left[\begin{array}{ccc} K_{11} & 
K^1_{12}& K^1_{13}\\
                          K^1_{21} & K^1_{22} & K^1_{23}\\
			  K^1_{31} & K^1_{32} & K^1_{33}
	\end{array}\right]
\]
and 
\[
YZ=\left[\begin{array}{cc} Y_1Z_1 & Y_1z_1\\
                          z_1^* & Z_{33}
	\end{array}\right]=\left[\begin{array}{ccc} 2 & 0& 0\\
                          0 & L^1_{22} & L^1_{23}\\
			  0 & L^1_{32} & L^1_{33}
	\end{array}\right]
\]
with the block decompositions of the matrices in $B(\iH_1\oplus \iK_a 
\oplus \iK_b)$ on the right-hand-sides. This implies that
$Z_1$ commutes with both $X_1$ and $Y_1$, $K^1_{33}=Z_{33}=L^1_{33}$ and
\begin{equation}\label{eq:xyz}
X_1z_1=Y_1z_1=z_1=\left[\begin{array}{c} 0\\
                          L^1_{23}
	\end{array}\right]
\end{equation}
and we are in a similar situation as before. (Compare with the relations
(\ref{E:has1}), (\ref{eq:K1}) and (\ref{eq:XYz}).) We also get 
$K^1_{12}L^1_{23}=0$
exactly as before. Note that now we can write
\[
K=\left[\begin{array}{cccc} K_{11} & K^1_{12}& 0 & 0\\
                          K^1_{21} & K^1_{22} & L^1_{23}& 0\\
			  0 & L^1_{32} & L^1_{33} & L_{23}\\
			  0 & 0 & L_{32} & L_{33}\end{array}\right]
\]
Again, if $L^1_{23}=0$ or if the range of $L^1_{23}$ is $\iK_a$, then 
the matrix $K$ is block diagonal. If this condition does not hold, then
the above procedure can be repeated: we decompose the subspace $\iK_{a}=
 \iK^2_{a}\oplus\iK^2_{b}$, where $\iK^2_{b}$ is the range of $L^1_{23}$, 
and write $X_1$, $Y_1$ in the block-diagonal form, using  (\ref{eq:xyz}), etc. 

After repeating this procedure $n$-times, we get the matrix $K$ in the form
\[
K=\left[\begin{array}{ccccc} K_{11} & K^n_{12}& 0 & 0 & 0\\
                          K^n_{21} & K^n_{22} & L^n_{23}&0& 0\\
			  0 & L^n_{32} & L^n_{33} &L^{n'}_{34}& 0\\
                          0 & 0 & L^{n'}_{43} &  L^{n'}_{44}& L_{23}\\
			  0&0&0 &L_{32} & L_{33}
			  \end{array}\right]
\]
in $B(\iH_1\oplus \iK^n_{a}\oplus \iK^{n}_{b}\oplus
\iK^{n'}_{b}\oplus \iH_3)$, where $\iK^{n'}_{b}=\oplus_{k=1}^{n-1} 
\iK^k_{b}$, $\iK^1_b\equiv \iK_b$ and $K^n_{12}L^n_{23}=0$. 
Since $\iH_2$ is finite-dimensional, 
there must be some $n$, such that the matrix $L^n_{23}$ is either 0 
or has range  $\iK^n_a$. In both cases, the  matrix $K$ has a block diagonal
form, and so does the matrix $A$. \qed

The CCR Markov triplets have some similarity to Markov states on a  product
algebra $M_n(\bbbc) \ot (M_u(\bbbc)\ot M_t(\bbbc)) \ot M_n(\bbbc)$ ($n=ut$).
If $\omega_1$ is a state on $M_n(\bbbc) \ot M_u(\bbbc)$ and 
$\omega_2$ is a state on $M_t(\bbbc) \ot M_n(\bbbc)$, then $\omega_1
\ot \omega_2$ is Markovian, but there are other Markov states, however
they are constructed essentially by this idea \cite{HA}.
 
\section{Connection to classical Gaussians}

Markov triplets in the classical Gaussian case were studied in \cite{APD}. 
The present non-commutative situation has some relation to the classical 
Gaussian.

\begin{lemma}
Let $e_1,e_2,\dots, e_k$ be linearly independent unit vectors in $\iH$
such that $\<e_i,e_j\>$ is real, $1 \le i \le j \le k$.
Then the Weyl unitaries $W(te_j)=\exp (t \im B(e_j))$ commute. With 
respect to a quasi-free state (\ref{E:Fstate2}), the joint distribution 
of the field operators $B(e_1), B(e_2), \dots, B(e_k)$ is Gaussian.

Assume that $f_1, f_2,\dots, f_k$ are orthonormal vectors and $Sf_i=e_i$
for a linear mapping $S$. The covariance is the matrix of the linear 
operator $S^*(I+2A)S$ in the basis $f_1,f_2, \dots, f_k$. 
\end{lemma}

\proof
The characteristic function of the joint distribution is
\begin{eqnarray*}
(t_1,t_2,\dots,t_j)&\mapsto& \omega_A(\exp (\im t_1 B_1) \exp (\im t_2 B_2)
\dots \exp (\im t_k B_k) \cr \cr  &=&
\omega_A (W(t_1 e_1+t_2 e_2+\dots +t_k e_k))\cr \cr &=&
\exp \Big( -\fel (\sum_{i,j} t_it_j \< e_i, (I+2A)e_j\>)\Big)
\cr \cr &=&
\exp \Big( -\fel (\sum_{i,j} t_it_j \< Sf_i, (I+2A)Sf_j\>)\Big).
\end{eqnarray*}
This gives the result. \qed

Next we assume that $\iH=\iH_1\osum \iH_2 \osum \iH_3$ and assume that
$\dim \iH_i=k$ $(1 \le i \le 3)$. Choose pairwise orthogonal unit
vectors $f_j$, $1 \le j \le 3k$ such that 
$$
f_{(i-1)k+r} \in \iH_i, \qquad 1 \le i \le 3, \quad 0 \le r \le k-1  
$$
and unit vectors $e_j$, $1 \le j \le 3k$ such that 
\begin{equation}\label{cl1}
e_{(i-1)k+r} \in \iH_i, \qquad 1 \le i \le 3, \quad 0 \le r \le k-1  
\end{equation}
and 
\begin{equation}\label{cl2}
\<e_t , e_u\> \mbox{\ is\ real\ for\ any\ }1 \le t,u \le 3k.
\end{equation} 
There is an invertible  block diagonal matrix $S=\Diag(S_1, S_2, S_3)$ such 
that $S f_j=e_j$, $1 \le j \le 3k$. 

The Weyl unitaries $W(te_j)=\exp (t \im B(e_j))$ commute. The joint 
distribution of the field operators  $B(e_j)$ is Gaussian with  covariance
block matrix $S^*(I+2A)S$. It follows from  \cite{APD} that the classical 
(multi-valued) Gaussian triplet 
$$
(B(e_1),\dots, B(e_k)),\quad
(B(e_{k+1}),\dots,B(e_{2k})), \quad
(B(e_{2k+1}),\dots,B(e_{3k}))
$$ 
is Markovian if and only if 
$$
[S^*(I+2A)S]_{13}=[S^*(I+2A)S]_{12} [S^*(I+2A)S]_{22}^{-1} 
[S^*(I+2A)S]_{23}. 
$$
Since
$$
[S^*(I+2A)S]_{13}=S_1^*(I+2A)_{13}S_3
$$
and
$$
[S^*(I+2A)S]_{12} [S^*(I+2A)S]_{22}^{-1} [S^*(I+2A)S]_{23}=
S_1^*(I+2A)_{12}(I+2A)_{22}^{-1}(I+2A)_{23}S_3,
$$
the matrix $S$ can be removed from the condition and we have the
equivalent form $(I+2A)_{13}=(I+2A)_{12}(I+2A)_{22}^{-1}(I+2A)_{23}$
which means that $(1,3)$ element of $(I+2A)^{-1}$ is 0. 
If the quasi-free state induced by $A$ gives a Markov triplet, then  
$A$ is the form of (\ref{E:trivi}) due to Theorem \ref{thm:markov}. In particular,
$(A^{-1})_{13}=0$ and reference to \cite{APD} gives the following result.

\begin{thm}
Let $\omega$ be a quasi-free state on $\CCR(\iH_1\osum \iH_2 \osum \iH_3)$
which is Markovian with respect to the decomposition $\iH_1\osum \iH_2 \osum \iH_3$.
Assume that $e_j$, $1 \le j \le 3k$ are unit vectors such that (\ref{cl1}) and
(\ref{cl2}) hold. Then the classical (multi-valued) Gaussian triplet 
$$
(B(e_1),\dots, B(e_k)),\quad
(B(e_{k+1},\dots,B(e_{2k})), \quad
(B(e_{2k+1},\dots,B(e_{3k}))
$$ 
is Markovian, moreover,
$$
(B(e_1),\dots, B(e_k))\quad \mbox{and} \quad
(B(e_{2k+1},\dots,B(e_{3k}))
$$ 
are independent.
\end{thm}

In a final remark we compare the Markov condition for the classical multivariate
Gaussian triplet with the CCR case. The classical condition is $A_{12}A_{22}^{-1}
A_{23}=A_{13}$. The CCR condition can be formulated as $A_{12}f(A_{22})A_{23}=A_{13}$
with any continuous function $f$. (This implies immediately that $A_{13}=0$.)
Therefore, the CCR condition is much more restrictive.

\end{document}